\documentclass{vgtc}                          

\graphicspath{{figures/}{./}} 

\usepackage{times}                     

\usepackage{mathptmx}                  

\onlineid{9} 

\vgtccategory{Research}

\vgtcinsertpkg

\usepackage{comment}
\usepackage{textcomp}
\usepackage{gensymb}

\title{Moving Beyond the Flat Screen: Opportunities and Challenges of Extended Reality for Studying Human Brain and Behavior} 

\author{Tom Bullock\textsuperscript{1,2,3}
\and Radha Kumaran\textsuperscript{4}
\and You-Jin Kim\textsuperscript{5}
\and Emily Machniak\textsuperscript{6}
\and Tobias Höllerer\textsuperscript{2,7}
\and Barry Giesbrecht\textsuperscript{1,2}}

\affiliation{
\vspace*{-1.5ex} 
\textsuperscript{1} \scriptsize{Department of Psychological and Brain Sciences, University of California, Santa Barbara, Santa Barbara, CA, United States.}\\
\textsuperscript{2} \scriptsize{Institute for Collaborative Biotechnologies, University of California, Santa Barbara, Santa Barbara, CA, United States.}\\
\textsuperscript{3} \scriptsize{Department of Social and Psychological Sciences, University of Huddersfield, Huddersfield, United Kingdom}\\
\textsuperscript{4} \scriptsize{Saarland University, Saarland Informatics Campus, Saarbrücken, Germany.}\\
\textsuperscript{5} \scriptsize{School of Computing, College of Engineering, University of Nebraska--Lincoln, Lincoln, NE, United States.}\\
\textsuperscript{6} \scriptsize{Department of Human Systems Engineering, Arizona State University, Mesa, AZ, United States.}\\
\textsuperscript{7} \scriptsize{Department of Computer Science, University of California, Santa Barbara, Santa Barbara, CA, United States.}\\
\vspace*{-3ex} 
}

\teaser{
  \centering
  \includegraphics[width= .88\linewidth]{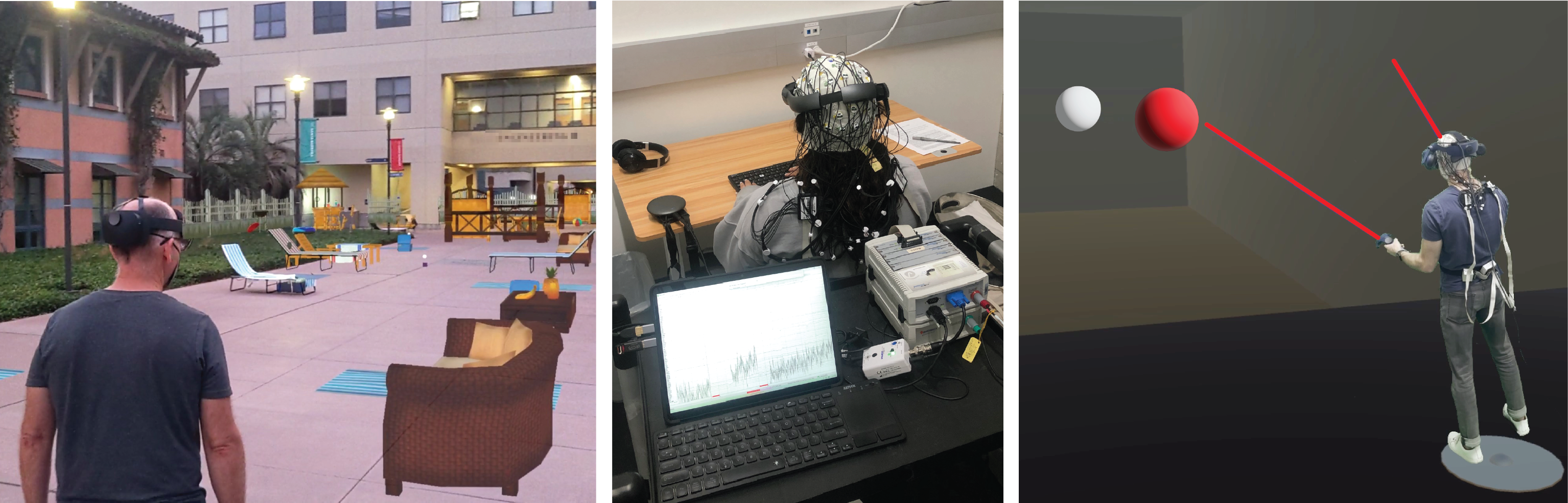}
  \setlength{\abovecaptionskip}{0pt}   
  \caption{Representative scenes from the three lines of our work discussed in this paper. Left: wide-area visual search. Participants use a head-worn AR display to search a large outdoor environment for virtual targets under different lighting conditions~\cite{kimInvestigatingSearchPhysical2022}. Center: semantic processing in AR. Seated participants view virtual objects and words in augmented reality while EEG is recorded and used to probe neural markers of semantic processing~\cite{bullock_electrophysiological_2025}. Right: spatial attention to dynamic stimuli. A participant, shown superimposed into the virtual environment, strikes incoming target spheres with a pair of virtual sabers anchored to HTC Vive controllers while EEG is recorded and used to decode the focus of spatial attention~\cite{bullockSABERSpatialAttention2026}. }
  \label{fig:Figure1_Teaser} 
}

\abstract{
Much of what we know about human perception and cognition comes from austere experimental settings: static stimuli presented to sedentary observers on flat monitors, complex responses reduced to simple button presses, and brain activity recorded from immobilized heads using EEG or fMRI. These highly controlled paradigms have provided remarkable insights, but capture a small and often unusual slice of behavior, raising questions about how well the findings generalize to the real world. Extended reality (XR), spanning both immersive virtual reality (VR) and augmented reality (AR) that overlays digital content on the real world, offers a way beyond these constraints, supporting controlled, reproducible experiments in which participants actively engage with three-dimensional tasks. Drawing on a growing body of VR and AR studies of attention, memory, and language---our own and others', some involving EEG---we outline what XR makes newly possible for cognitive neuroscience and reflect on the practical and conceptual obstacles that remain. The opportunities are substantial, but realizing them requires solving practical problems of hardware, timing, and measurement, and being realistic about what a virtual or augmented world can and cannot stand in for.
}


\makeatletter
\renewcommand{\copyrightspace}{%
   \renewcommand{\thefootnote}{}%
   \footnotetext[0]{\textit{This is a preprint version of a workshop paper presented at the \mbox{NeuroXR} 2026 Workshop at IEEE ISMAR 2026. The final version appears in the 2026 IEEE International Symposium on Mixed and Augmented Reality Adjunct (ISMAR-Adjunct). For citation, please refer to the published version. This work was initially made available on the author's personal website [yujnkm.com] in September 2026, and was subsequently uploaded to arXiv for broader accessibility.}}%
   \renewcommand{\thefootnote}{\arabic{footnote}}}
\makeatother

\begin{document}

\firstsection{BACKGROUND}
\maketitle 

We write as an interdisciplinary team of cognitive neuroscientists and computer scientists, drawn together by a shared interest in the human brain, complex behaviors, and how to study them. Much of our understanding about perception and cognition from experimental psychology and neuroscience rests on a particular methodological recipe. Carefully controlled static stimuli are presented on flat computer screens to undergraduate participants, who hold their eyes still. Behavior is measured with simple key presses, and sometimes brain activity is recorded with electroencephalography (EEG) or functional magnetic resonance imaging (fMRI). The rigor of this approach is its great virtue and the source of much of what we know about the mind. This austerity is a deliberate design choice, as stripping away the messy confounds of the natural environment allows cognitive mechanisms to be isolated and measured with precision. But it is also a strange way to study behavior. Real perception unfolds while people walk, turn their heads, reach for things, and sample a cluttered three-dimensional world with their eyes. Almost none of these features of naturalistic behaviors are present in the standard methodology. This therefore raises concerns, not with the validity of the theories that this carefully controlled lab work has established, but with how far findings obtained under such constrained laboratory conditions generalize to the natural world.

For the past several years, we have collaborated across our disciplines to move beyond these traditional methods, using XR to study brain and behavior in richer, more naturalistic tasks and environments — and to study cognition during XR use itself, to better understand the benefits and side effects the technology brings (Fig.~\ref{fig:Figure1_Teaser}). These collaborations have been remarkably productive, precisely because no single discipline holds all the necessary skills: designing a rigorous, theoretically motivated attention or memory paradigm, building a performant XR environment, recording clean mobile EEG, and modeling the resulting data, each demand different expertise. The opportunities are considerable, but so are the obstacles. Here, we outline what we view as the main opportunities, drawing on examples from across the field: studies of perception and memory in
VR~\cite{schoneRealityVirtualReality2023,draschkowMultipleSpatialFrames2022,kiskerVisualInformationProcessing2025}; visual search, both in confined displays~\cite{chiossiSearchingRealitiesInvestigating2024} and in our own work across wide, naturalistic areas ~\cite{kimInvestigatingSearchPhysical2022,kumaranImpactNavigationAids2023,kimGoARAttention2025,kumaranSceneAwarenessUsing2025}; dynamic object tracking in VR with simultaneous EEG~\cite{bullockSABERSpatialAttention2026}; and efforts to transfer well-established laboratory paradigms into immersive settings, testing whether their classic neural signatures survive ~\cite{sagehornElectrophysiologicalCorrelatesFace2024a,bullock_electrophysiological_2025}. We then turn to practical and conceptual challenges that must be overcome.

\section{OPPORTUNITIES}

\subsection{Ambulatory testing} 
A key advantage of XR is that stimuli can be presented to participants while they are in motion, making it possible to study how people navigate large spaces and search for goal-relevant items as they move through the environment. We have used the HoloLens 2 to examine wide-area search outdoors ~\cite{kimInvestigatingSearchPhysical2022,kumaranImpactNavigationAids2023} and indoors ~\cite{kimGoARAttention2025,kumaranSceneAwarenessUsing2025}~(Fig.~\ref{fig:Figure1_Teaser}, left). This work taught us how attention operates when people must attend to both physical and digital worlds at once, and how this affects their situational awareness, with direct implications for the design of AR hardware and of applications intended to be used on the go. Others have combined VR with mobile EEG to examine the cortical correlates of naturalistic navigation ~\cite{delauxMobileBrainBody2021}.

XR also makes it possible to study how changes in physiological state shape performance. In our own work, physical exertion has been shown to measurably alter attention and working memory processes ~\cite{garrett_systematic_2024,giesbrecht_physically_2025,garrett_tracking_2021}, and mobile AR allows these effects to be measured while participants move through the environment~\cite{cao_walking_2019}. 

\subsection{Moving beyond the button press} 
Cognitive psychology and neuroscience have long treated perception and action as separable stages, collapsing a rich perceptual decision onto a single key press and treating the motor response as a readout of a cognitive process presumed complete before movement begins. Yet, a growing body of work argues that attention and action are tightly coupled: what we select is shaped by, and unfolds concurrently with, the movements we are preparing~\cite{songAbandoningModifyingOne2017,vanedeConcurrentVisualMotor2019}. Recent studies show that visual and motor selection proceed together and scale jointly with the reliability of the cue guiding behavior~\cite{wangWangSiSiProactiveVisualMotor2026}, and that a planned action can warp the representation of visual features themselves~\cite{trentin_action_2024}. This renewed interest in perception--action coupling is precisely what traditional button-press paradigms are not well-equipped to capture. XR offers a natural remedy: platforms continuously capture controller or hand position, head pose, and gaze, yielding rich and objective measures of behavior as it unfolds and letting us study cognition as it guides action, rather than as a process inferred from a single button press.

\subsection{Dynamic 3D stimuli}
Research on visual spatial attention and working memory remains dominated by static, 2D displays, even though everyday perception constantly engages objects moving in 3D. Studies on multiple-object tracking are an important exception~\cite{meyerhoffStudyingVisualAttention2017}, but with several targets present at once, it is hard to isolate the brain's response to any individual object. XR makes controlled, dynamic, three-dimensional stimuli routine, permitting investigations of how attention is sustained and shifted among moving objects at different depths, closer to the demands the system evolved to meet. In our own recent work we adapted the popular VR game Beat Saber into a rigorous paradigm: participants intercept individual targets that loom toward them while we record their EEG, and multivariate analyses reconstruct the focus of spatial attention as each target moves~\cite{bullockSABERSpatialAttention2026}~(Fig.~\ref{fig:Figure1_Teaser}, right). We validated this platform, which offers a promising testbed for future work on dynamic attention and its coupling with action. Immersion also lets us probe spatial working memory in new ways: when items surround an observer in three dimensions, rather than sitting on a flat screen, working memory can exploit multiple spatial reference frames at once~\cite{draschkowMultipleSpatialFrames2022}.

\subsection{Immersion reshapes neural responses}
Presence and immersion are not only a matter of engagement: when a participant is immersed in a task or environment in the first person, rather than watching it on a flat screen, a growing body of work shows neural responses can diverge from those a traditional monitor elicits. A particularly compelling demonstration comes from a height exposure study, where participants were raised to 33~m in a fire truck's aerial basket, viewed a matched 3D-360$^\circ$ video of the ascent in VR, or watched a 2D version on a screen in the lab. EEG markers of vigilance and arousal, and heart-rate variability, were nearly indistinguishable between real and virtual exposure, yet both diverged sharply from the flat-screen~\cite{schoneRealityVirtualReality2023}. The same contrast appears in visual and memory processing: presenting objects in VR rather than on a monitor shifts object-processing signatures toward real-world patterns and reduces cognitive load~\cite{kiskerVisualInformationProcessing2025}. Even the N170 face-selectivity effect (stronger response to faces than other objects), a hallmark of monitor-based studies, is attenuated under realistic VR viewing, where later components carry the discrimination between faces and objects~\cite{sagehornElectrophysiologicalCorrelatesFace2024a}. This makes immersion a manipulable variable in its own right, letting us test whether established neural signals survive beyond the 2D monitor.

For the experimenter, this brings both opportunity and risk. Where immersive presentation elicits stronger or more naturalistic responses, it may raise a recording's signal to noise ratio, making effects less likely to be masked by noise. However, caution is advised, as a paradigm validated on a 2D screen may not behave identically in an immersive one: the signal carrying an effect may migrate to another component, change in amplitude, or shift in time.

\subsection{Sensor integration}
Eye-tracking is now part of many headsets, turning gaze into a routine measure rather than a specialized add-on. Gaze is at once behavior, indexing where attention is deployed, and a signal to be modeled. It is also a potential contaminant that may distort other possible signals of interest. For example, eye movements can distort concurrent EEG measures of spatial attention, so they must be accounted for, not ignored~\cite{bullockEyeMovementsDisrupt2023}. Having eye-tracking available by default reshapes the suite of scientific questions that can be asked, although tracking quality can be an issue. Modern headsets track head orientation and movement and routinely collect ancillary sensor data, including accelerometer and gyroscope readings, facial expressions, and related biometric signals. Combined with EEG, these streams support rich multimodal readouts of a user's cognitive state. This wealth of sensitive data about the individual, however, also raises sensitive ethical issues (see Sec.~\ref{sec:ethics}).

\section{CHALLENGES}

\subsection{Building the experiment} 
\label{sec:building}
For a cognitive psychologist or neuroscientist, the single largest barrier is often simply building the experiment. Many of us have limited formal programming experience: we can implement simple paradigms in MATLAB with Psychtoolbox-3 or Python with PsychoPy, but a game engine such as Unity (programming in C\#) is a substantial step up. Game engines are powerful, but they were built to render rich graphics, not to deliver the millisecond stimulus timing, precise event logging, and tight experimental control that psychophysics requires. This is changing quickly, however: research-oriented open-source toolboxes such as simpleXR (\url{https://github.com/simpleOmnia/sXR}), commercial platforms such as Silico Labs (\url{https://www.silicolabs.ca/}), and LLM-based coding assistants have substantially lowered the difficulty and time cost of building XR experiments, making them more accessible to psychologists and neuroscientists. Some problems, though, remain stubbornly technical, as discussed next.

\subsection{Designing large-scale augmented environments} 
\label{sec:designing}
Running rigorous user studies in wide-area environments, such as one of our own examples encompassing 1,456 $m^2$~\cite{kimInvestigatingSearchPhysical2022}, demands meticulous spatial planning and accurate asset tracking. Beyond item placement, it requires occlusion layers so that virtual content is masked naturally by physical structures, and an accurate yet lightweight digital twin that mirrors the orientation and scale of the venue. Scale fidelity is critical, as a 0.1 scale error in Unity becomes a 10-centimeter error on the ground. Meeting these constraints requires a dedicated capture pipeline, such as ours which combines satellite imagery with localized laser-range measurements to achieve exact scale matching, which we document elsewhere~\cite{kimInvestigatingSearchPhysical2022}.

\subsection{Timing and synchronization}
\label{sec:timing}
Many questions in cognitive neuroscience depend on precise stimulus timing and event-locking of behavior and brain activity, yet XR devices are rarely built with this in mind. In our own combined AR/EEG study of semantic processing ~\cite{bullock_electrophysiological_2025}~(Fig.~\ref{fig:Figure1_Teaser}, center), we needed to recover event-related potentials which required precise timing, but the Magic Leap headset exposed no trigger output that could be aligned to the EEG. Our workaround was to emit a tone at each critical event, convert it to an electrical pulse and record it alongside the EEG, recovering the markers offline. This worked but was cumbersome and required extra hardware (documented in~\cite{bullock_electrophysiological_2025}). More general solutions are emerging, however. The Lab Streaming Layer (LSL) provides a common clock for synchronizing across data streams, and is widely used across labs, though its integration with self-contained AR headsets remains uncertain.

\subsection{Headsets}
\label{sec:headsets}
Many cognitive psychology and neuroscience labs, our own included, still rely on the aging HTC Vive Pro Eye for VR research. It is dated and discontinued, yet its wireless-adapter support remains valuable for ambulatory studies, where a tether would constrain movement. An ideal replacement would be natively wireless and lightweight, with precise headset and controller tracking and high-quality eye tracking, a tall order that devices such as the Samsung Galaxy XR, the forthcoming Pico Project Swan, and Valve Steam Frame may soon meet. Moving to AR brings different challenges. The commercial market for standalone optical-see-through displays is transitioning from capable but bulky headsets nearing end of support (HoloLens 2, Magic Leap 2) toward unobtrusive, continuous-wear options that are promised but not yet widely available (e.g., Meta Ray-Ban Display, Snap Specs), with some consumer glasses aimed at stabilized media consumption (e.g., Viture Luma Ultra, XREAL One Pro) not yet readily adapted for research. 

Outdoors, bright sunlight frequently causes headsets such as the HoloLens~2 to lose wide-area tracking, so that reliable operation is possible only in shade, meaning that the position of the sun constrains scheduling. Displays are similarly bounded: current optical see-through panels struggle at both extremes of ambient light, washing out under bright sunlight and glowing too brightly in near-darkness. On the software side, development still depends heavily on the Mixed Reality Toolkit (MRTK3) for both the HoloLens 2 and the Magic Leap 2: relying on native OpenXR is not yet feasible where experiments require legacy features and spatial anchors.

\subsection{Tracking}
\label{sec:tracking}
Accurate spatial computing depends on continuous, precise tracking of the head, eyes, and body. Current systems are usable but imperfect: inside-out tracking degrades in feature-poor, poorly lit, or highly repetitive environments, and tracking over large or outdoor spaces remains a major hurdle. Add the difficulty of tracking hand position wherever the hands are operating, and inferring leg position from available sensors, and a perfect solution looks far off.

These problems are not hypothetical: in our own wide-area outdoor AR studies~\cite{kimInvestigatingSearchPhysical2022,kumaranImpactNavigationAids2023}, we found that momentary tracking loss can shift the virtual world, disorienting participants and disrupting tasks. One mitigation, Azure Spatial Anchors (since deprecated), pinned virtual content to fixed points in the environment. Effective as this was, anchors had to be manually re-aligned each time the application restarted, adding considerable overhead to sessions.

\subsection{Ethical considerations}
\label{sec:ethics}
The same wealth of sensor data that makes XR so powerful also makes it invasive. Modern headsets can continuously capture gaze, facial expressions, gait and posture, and even neural activity. These signals can be mined to infer potentially sensitive attributes that a user may not intend to disclose, such as moment-to-moment attention, affective state and even identity. Recent work on cognitive biometrics and mental privacy warns that such inferences fall beyond what current data-protection protocols were designed to cover~\cite{mageeNeuralDataCognitive2024a}. One response is a multi-layered privacy-permission framework for XR that lets users govern not just which data are recorded but which inferences may be drawn from them~\cite{mansourMultiLayeredPrivacyPermission2025}. This complicates open science, since the rich raw recordings needed for reproducible data sharing can re-identify participants, undermining anonymity.

\subsection{XR is not reality}
However immersive XR becomes, it remains a simulation, not the real world itself. At the sensory level, rendered depth is not real depth: vergence and accommodation cues conflict, scale can be distorted, and the sensory consequences of action are only approximated. These limits matter for any study that hinges on fine perceptual or sensorimotor fidelity.

A subtler and arguably more important issue concerns behavior rather than perception. It is tempting to assume that a more immersive, realistic environment yields more natural behavior. But as Yartsev argues in the context of animal neuroscience, natural behavior is defined by what an organism does---whether a task engages the actions it evolved to perform and whether the environment affords them---not by where it is~\cite{yartsevNaturalBehaviorWhat2026}. The same caution applies to human XR, perhaps with greater force, since a headset can render a compellingly realistic world while the task it imposes bears little relation to ordinary behavior. Realism is not sufficient for studying natural behavior and can even work against it, creating an illusion of ecological relevance or eliciting outright unnatural behavior. This reflects a more basic methodological trade-off between internal and external validity. The very features that allow for external validity, such as wide-area outdoors testing, free movement and rich 3D scenes, are the ones that can erode internal validity by introducing the timing, synchronization and tracking problems cataloged above as uncontrolled variance. This risk is sharpened by evidence that immersive VR can approximate real experience at the psychophysiological level~\cite{schoneRealityVirtualReality2023}: because a virtual environment can feel and measure like the real world, its ecological validity is easily overstated. The burden thus remains on the experimenter to design tasks that afford genuinely natural actions, and how far XR findings generalize to the unconstrained world remains an open empirical question.

\section{CONCLUSION}
XR provides cognitive psychology and neuroscience with an enticing combination: experimental control together with the freedom to study people as they move, look, and act. VR has served this purpose for some time, but AR is relatively new and promising. As head-worn AR approaches a form factor that people might plausibly wear in public, with devices such as Snap's Specs coming close to that mark, it becomes pressing to understand how AR shapes attention, memory, and situational awareness before the technology is widely adopted. At the same time, from the perspective of studying human behavior, we must be honest about XR's limits: it is a powerful, controllable approximation of the world, not the world itself. Realizing the promise of XR will take exactly the cross-disciplinary exchange this workshop convenes, making it an ideal forum to weigh these opportunities against their challenges.

Before committing to an XR study, a few checks distilled from the challenges above can save the experimenter considerable time. The build must be feasible, since the expertise and effort to construct a paradigm, especially in large or complex environments, is easily underestimated (Secs.~\ref{sec:building} and ~\ref{sec:designing}); timing must be precise wherever events are locked to physiological signals such as EEG (Sec.~\ref{sec:timing}); the headset must suit the task, wireless if ambulatory and with good eye-tracking when needed (Sec.~\ref{sec:headsets}); outdoor work brings added obstacles, from bright sunlight to unreliable wide-area tracking (Sec.~\ref{sec:tracking}); and consent should cover the inferences the sensors permit, not merely the data one intends to analyze (Sec.~\ref{sec:ethics}).

\acknowledgments{Our work was supported by the U.S. ARO (W911NF-09-D-0001, W911NF-19-D-0001 [ICB], W911NF-19-2-0026), Army Soldier Center MASTR-E program (W911NF-19-F-0018), ONR (N00174-19-1-0024; N00014-19-1-2553, -20-1-2719, -23-1-2118), NSF (IIS-1911230, IIS-2211784), and DARPA (HR001124C0409).}

\bibliographystyle{abbrv-doi}


\bibliography{references}

\end{document}